# Frustrated phonon with charge density wave in vanadium Kagome metal

Seung-Phil Heo[1,2,3], Choongjae Won[4], Heemin Lee[1,2,3], Hanbyul Kim[5], Eunyoung Park[1,2,3], Sung Yun Lee[1,2,3], Junha Hwang[1,2,3], Hyeongi Choi[6], Sang-Youn Park[6], Byungjune Lee[1,3,4], Woo-Suk Noh[4,6], Hoyoung Jang[3,6], Jae-Hoon Park[1,3,4], Dongbin Shin[5,7*], Changyong Song[1,2,3*]

*[1]Department of Physics, POSTECH, Pohang 37673, Korea. [2]Center for Ultrafast Science on Quantum Matter, Max Planck POSTECH/Korea Research Initiative, Pohang 37673, Korea. [3]Photon Science Center, POSTECH 37673, Pohang 37673, Korea. [4]Center for Complex Phase Materials, Max Planck POSTECH/Korea Research Initiative, Pohang 37673, Korea. [5]Department of Physics and Photon Science, Gwangju Institute of Science and Technology, Gwangju 61005, Korea. [6]Pohang Accelerator Laboratory, POSTECH, Pohang 37673, Korea. [7]Max Planck Institute for the Structure and Dynamics of Matter, Hamburg 22761, Germany.*

**The formation of a star-of-David CDW superstructure, resulting from the coordinated displacements of vanadium ions on a corner-sharing triangular lattice, has garnered significant attention to comprehend the influence of electron–phonon interaction within geometrically intricate lattice of Kagome metals, specifically $AV_3Sb_5$ (where A represents K, Rb, or Cs). However, understanding of the underlying mechanism behind CDW formation, coupled with symmetry-protected lattice vibrations, remains elusive. Here, from femtosecond time-resolved X-ray scattering experiments, we reveal that the phonon mode, associated with Cs ions' out-of-plane motion, becomes frustrated in the CDW phase. Furthermore, we observed the photoinduced emergence of a metastable CDW phase, facilitated by alleviating the frustration. By not only elucidating the longstanding puzzle surrounding the intervention of phonons but introducing the**

**phononic frustration, this research offers fresh insights into the competition between phonons and periodic lattice distortions, a phenomenon widespread in other correlated quantum materials including layered high-$T_C$ superconductors.**

Crystals with unique ionic arrangements and strong electronic correlations serve as a fertile ground for the emergence of exotic phases, as evidenced by the coexistence of charge density wave (CDW) and superconductivity in vanadium (V) Kagome metals of $AV_3Sb_5$ (A for K, Rb, and Cs). Exotic phases involving CDW and superconductivity (SC), arising from intricate electronic structures within geometrically frustrated ionic arrangements, have spurred active investigation into the Kagome metal, $AV_3Sb_5$ (Fig. 1a)[1,2]. Experimental findings suggest that mechanisms driving CDW formation extend beyond Peierls distortion that is facilitated by enhanced electron–phonon interactions through nested Fermi surfaces. This has led to new hypotheses regarding electron–hole pairing mediated by phonons[3-6]. The van Hove singularity of V *d*-electrons near the Fermi level has been proposed as a pivotal factor in CDW instability[7,8]. However, the role of phonons remains ambiguous, as phonon softening, a characteristic of CDW instability, has not been observed[5,9]. Additionally, the CDW and SC phases exhibit sensitivity to different choices of alkali metal ions in the Kagome system[10-14], despite the limited presence of electron population from the cations near the Fermi level[7,8]. These findings suggest that the electronic potential energy surface may undergo dramatic changes due to cation movement, giving rise to such diverse phases. While there have been several ultrafast optical and ARPES studies on CDW phases[15-17], further research into the microscopic CDW orderings in excited states remains necessary.

We conducted femtosecond (fs) time-resolved X-ray scattering experiments on $CsV_3Sb_5$ to investigate CDW instabilities within a finely balanced potential energy landscape, induced by photoexciting electrons near the Fermi level of the range of ~ 1.5 eV matching with the photon energy of fs-infrared (IR) laser pulses. Mostly electrons in V and Sb ions' orbitals are involved in this photoexcitation process[5,18]. X-ray scattering experiments were conducted utilising the Pohang Accelerator Laboratory X-ray Free Electron Laser (PAL-XFEL) and the Pohang Light Source (PLS). The emergence of superstructures below the CDW transition temperature of 96 K was confirmed, revealing the coexistent 2×2×1 and 2×2×2 CDWs (notated as $a \times b \times c$ in multiples of the unit cell) (Figs. 1a and b)[19]. Doubling of the unit cell in the *ab*-plane resulted in a 2×2 superstructure due to in-plane displacements of V ions, forming a star-of-David (SoD) or its inverse triangle-and-hexagon (TrH) pattern (Fig. 1a)[10,20,21]. Additionally, unit cell doubling along the *c*-axis was observed, establishing the 2×2×2 superstructure. Another lattice instability is required to induce this superstructure with the phonon mode ($L_2^-$) involving ions' *c*-directional motion attributed to driving this instability [9,15]. The 2×2×1 and 2×2×2 CDWs, corresponding to in-plane and out-of-plane distortion, respectively, were confirmed by differing peak widths and the temperature dependence of intensities at $Q_{CDW}^{2\times2\times1}$ = (0.5 0 Integer) and $Q_{CDW}^{2\times2\times2}$ = (0.5 0 Integer + 0.5). The order parameters are derived from the integrated intensities of the CDW satellites (Fig. 1b and Supplementary Note 1)[5,6,22].

The rapid redistribution of electrons near the Fermi level due to fs-IR photoexcitation disturbed these CDW orderings. Subsequently, the ultrafast reaction dynamics of the CDWs were investigated using femtosecond X-ray pulses from the PAL-XFEL (Fig. 1c and Method). The temporal evolution of peak intensity from the 2×2×2 CDW reflection exhibited an oscillation without full recovery until approximately 2 ns, whereas no oscillation was observed for the 2×2×1 CDW (Fig. 1d). These features were explicitly resolved by simulating exponential decay and recovery reactions (drawn with a solid line) (Supplementary Note 2). Both CDWs underwent rapid melting at approximately 100 fs, followed by a quick yet partial recovery of intensity in 300 fs. These timescales match with previously reported in CDW systems with electron–phonon interactions[17,23,24]. After the photoinduced melting, rapid reconstruction of the CDW within hundreds of femtoseconds was observed, but into a metastable state with reduced intensity involving an asymmetric vibration of Cs ions changed from the intact CDW's native $L_2^-$ mode. This metastable state, persisting for nanoseconds, is induced by photoinduced electronic perturbation rather than thermal disorder as noted by slow recovery of Bragg reflection in several hundred picoseconds (Supplementary Note 3)[25].

The oscillating intensity of the 2×2×2 CDW reflected a coherent phonon with a frequency of approximately 1.3 THz. This resembles aforementioned $L_2^-$ phonon but with a critical difference[9,15]. One should note that native $L_2^-$ phonon involves symmetric motions of two Cs ions moving upward while the other two move downward, exhibiting a specific symmetry. From the $3Q$ degenerate CDW structure of this system, this phonon mode can affect the 2×2×2 superstructure intensity but only for the one case out of three CDWs, as a minor contribution. (Supplementary Note 4)[15,20]. Hence, the observed oscillatory intensity suggests a fundamental modification of the native $L_2^-$ phonon, which induces intensity variations across all triply degenerate CDW states, caused by the fs-IR photoexcitation of conduction electrons. Consistent with this, the coherent phonons detected through CDW reflection exhibited an unusual phase delay of approximately $\pi$ (Supplementary Note 2)[26,27]. This phase delay indicates the presence of competing interactions that delay the prompt excitation of coherent phonons and distort the native phonon mode[28-30].

The emergence of strong diffuse signals along the *c*-direction around the 2×2×2 CDW reflection strongly supports the interpretation of competing interactions between the 2×2×2 CDW and the $L_2^-$ phonon mode (Fig. 2a). The first diffuse streak, which developed at 200 fs, was of the ordinary type resulting from CDW melting, with a reduced domain size of approximately 60% of the intact value (Fig. 2b). As the peak sharpened with CDW reconstruction, another strong diffuse streak appeared at around 400 fs, coinciding with the delayed onset of the coherent phonon (Fig. 2b). This transiently formed diffuse signal, observed at approximately 400 fs, exhibited a two-wing structure relative to the main CDW reflection.

These diffuse scattering peaks developed at positions with symmetric offsets ($\pm\ \delta q \sim 5\times10^{-3}$ $2\pi/c$) relative to the 2×2×2 CDW superstructure at 1.5 ($2\pi/c$), arising from disordered phase slip while relieving the frustrated phonon (Fig. 2b)[31-33]. The two-wing structure represents random period in the phase slip (Supplementary Note 5).

Other possibilities on the origin of the diffuse peaks have been attempted to include thermal diffuse scattering[34], order-disorder transitions[35], or critical scattering. Observed broad diffuse tail extending to inelastic excitation up to ~3 eV, and with symmetric contribution, is beyond ordinary quasiparticle excitations. The order-disorder transitions often involving in single broad diffuse scattering is not the current case. A possibility of critical scattering with two lengths scale, ordered region with long correlation along with disordered domains short correlation length, can be considered, which usually happen near the phase transition point[36]. We do not explicitly rule out this possibility but want to mention that such two length scale peaks were not observed in static temperature dependent measurement, not to support such interpretation. Yet, our phase-slip domain interpretation generally accounts for the observed intensity pattern.

Here, we present a schematic model describing the phase slip resulted from the asymmetric alterations: three-up (down) and one-down (up) deviated from the symmetric two-up and two-down displacements (Fig. 2c). The proposed structural model was consistent with the diffuse scattering pattern developed along the *c*-direction (Supplementary Note 5). One notes that increased disorder in phase slip domain size results in broadened width of the satellite diffuse peaks to bury the two-peak structure to a single diffusive peak alongside a sharp CDW reflection (Supplementary Note 5). However, this also reflects on a phase slip consistent with the present interpretation.

Further to confirm the modification of the native $L_2^-$ phonon mode after the photoexcitation, we have carried out time-dependent density functional theory (TD-DFT) calculations (Methods). The TD-DFT revealed that photoexcited electrons predominantly originate from V and Sb orbitals, as shown in density of occupied states (Fig. S6b). This photoexcitation disturbs the local electrostatic potential experienced by Cs atoms to promote site-dependent displacements (Supplementary Video 1). Accordingly, this photoinduced CDW breathing motion reshapes the potential energy to cause inverted displacement of one Cs ion, resulting in a three-to-one vibration (Fig. S6c). This inversion of the Cs-ion displacement was also corroborated by potential energy surface calculations, indicating that the system reaches a lower energy state with the Cs ions at the centre of TrH (or SoD) moving toward (out of) the V superstructure (Fig. 2e). The calculation confirmed an energy gain of 6 meV by displacing the Cs ions toward the TrH centre by 0.1 Å compared to the original position (Fig. 2e). In the native $L_2^-$ phonon mode, for the paired Cs ions, one ion was situated at the centre of the SoD (or TrH), whereas the other was not (Fig. 2d). Therefore, the native $L_2^-$ phonon mode with two-up and two-down vibrations was incompatible with the lower energy configuration, thus being frustrated in the CDW phase. With a significant perturbation of the potential energy landscape by redistributing electronic states in the V and Sb orbitals through fs-IR laser excitation, the CDW was transiently restructured. The phonon frustration became alleviated by inverting the movement of one Cs ion to align with the lower total energy configuration of the CDW phase. This resulted in asymmetric three-up (down) and one-down (up) vibrations, consistent with the observation of the oscillatory intensity of the 2×2×2 CDW reflection (Fig. 2d).

To comprehend the impact of photoexcited electrons on frustrated phonons and CDW instability systematically, we explored the fluence dependence of fs-IR pump laser (Fig. 3). A more pronounced reduction in peak intensities was consistently observed owing to increased CDW melting with higher laser fluence (Fig. 3a). The response times for melting and subsequent short recovery appeared relatively unaffected by variations in laser fluence ranging from 0.05 to 2.5 mJ $cm^{-2}$ (Figs. 3c and d). The melting time remained approximately 100 fs, displaying minimal dependence on laser fluence (Fig. 3c). This timescale is similarly noted for previously reported dynamics of nesting-mediated CDW systems[17,23,24]. Following this melting, there was a prompt intensity recovery, yet it failed to fully return to the intact state with the reopening of the CDW gap. This rapid recovery occurred at ~ 300 fs, consistent with both the 2×2×1 and 2×2×2 CDWs (Fig. 3d).

The fluence-dependent intensity of the 2×2×2 CDW exhibited intriguing characteristics warranting further investigation. Following the initial reduction upon melting, the intensity rebounded to its maximum value, showing a discernible dependence on laser fluence (Figs. 3a and b). For laser fluences exceeding approximately 0.5 mJ $cm^{-2}$, the intensity surpassed the intact state, denoted as 'underdamped CDW'. This transient CDW underdamping was similarly observed in other layered transition metal compounds exhibiting SoD CDW[37,38]. Subsequent to underdamping at around 400 fs, the intensity oscillated around a mean value lower than the intact state, without full recovery, persisting in a metastable state for several nanoseconds[15,16,39]. The intensity reduction in this metastable state increased with IR laser fluence until reaching approximately 1.4 mJ $cm^{-2}$, with a fluence dependence akin to that of the underdamped CDW intensity (Fig. 3b). Overall, the oscillation frequency of the modified coherent phonon remained largely unaffected by laser fluence, exhibiting only weak softening (1.5%) (Supplementary Note 8) [5,9,17].

We present a visual representation summarizing the overall photoinduced melting and metastable configuration of the 2×2×2 CDW (Fig. 4). The harmonic potential of the ions is schematically depicted for the CDW amplitude or mean ionic displacement (A). The potential energy of the intact 2×2×2 CDW exhibited a ground state with an initial CDW amplitude of $A_0$ (Fig. 4a)[40]. Fs-IR laser pumping shifted the potential energy surface (PES) to the ground state without a CDW (A = 0), indicating photoinduced gap closing (t ~ 0). With this transiently formed PES, the CDW melted to reach the ground state ($\omega t \sim \pi/2$)[41]. While adapting to this new PES, CDW underdamping occurred at $\omega t \sim \pi$ with the displacement, $A_{UD}$, further shifting the PES to a metastable ground state at $A_{MS}$. The new CDW then oscillated around the metastable ground state with an amplitude $A_{MS}$ smaller than $A_0$, as observed experimentally[15,16,39]. Corresponding atomic configurations are depicted (Fig. 4b). The intact CDW, characterised by two-up and two-down displacements of Cs ions relative to the V-dimerised CDW, demonstrated frustration. Upon femtosecond modification of the electron distribution by fs-IR photoexcitation, the symmetric (two-up and two-down) phonon mode, previously frustrated, was relieved, inducing three-up (down) and one-down (up) vibrations compatible with the potential energy surface.

In summary, we investigated the ultrafast dynamics of CDWs in the excited PES of $CsV_3Sb_5$ by transiently redistributing electrons near the Fermi levels using an fs-IR laser pulse. Through fs-IR photoexcitation, we induced a metastable CDW phase with a localised phonon mode, relieving the energetically frustrated two-up and two-down vibrations of Cs ions in the CDW state. This phonon mode, characterised by symmetric out-of-plane displacements of alkali ions, leads to energy loss directly coupled with the dimerization of V ions within the Kagome net, resulting in phonon frustration in the CDW phase. Restructuring the CDW using fs-IR laser photoexcitation relieves it by inducing a frustration-relieved localised phonon mode in a metastable CDW state. Our study unveiled the intricate coupling of phonons in this strongly correlated electronic system with exotic CDW phases by directly observing the modified asymmetric coherent phonons formed in a transiently disturbed potential energy configuration. It provides a concrete understanding on the mysterious roles of phonons in Kagome metals, emphasizing the crucial role of alkali ions in controlling CDW ordering[10-14]. While the phonon mode involving alkali ions' vertical motion emerges in the CDW state, noted with negative frequency in the phonon energy calculation[15], we found that this phonon mode becomes unstable in the CDW superstructure. The fs-IR photoexcitation mitigates this frustrated phonon to alter the vibration as three-up (down) and one-down (up) type, more compatible with the CDW structures. We also note that this frustration may defer immediate involvement of the $L_2^-$ phonon to induce the CDW, explaining the absence of phonon softening in this system[5,9].

We expect this frustrated phonon dynamics, being unrecognized so far, is rather widespread in emerging materials. Whilst no explicit relation of the observed frustrated phonon dynamics to the superconducting transition is drawn in this study, presence of a phonon mode becoming inactive, in layered high-$T_C$ superconductors and other layered materials, is discovered from theoretical investigations, which alludes the impact of phonon frustration in keeping the systems in the superconducting state[42]. This phonon frustration observed in this Kagome net can be direct evidence. This discovery suggests prompt investigation of similar dynamics in other correlated electronic systems and further shares common ground with understanding the energetics behind exotic charge ordering in layered high-$T_C$ superconductors and other two-dimensional dichalcogenide materials[43-47].

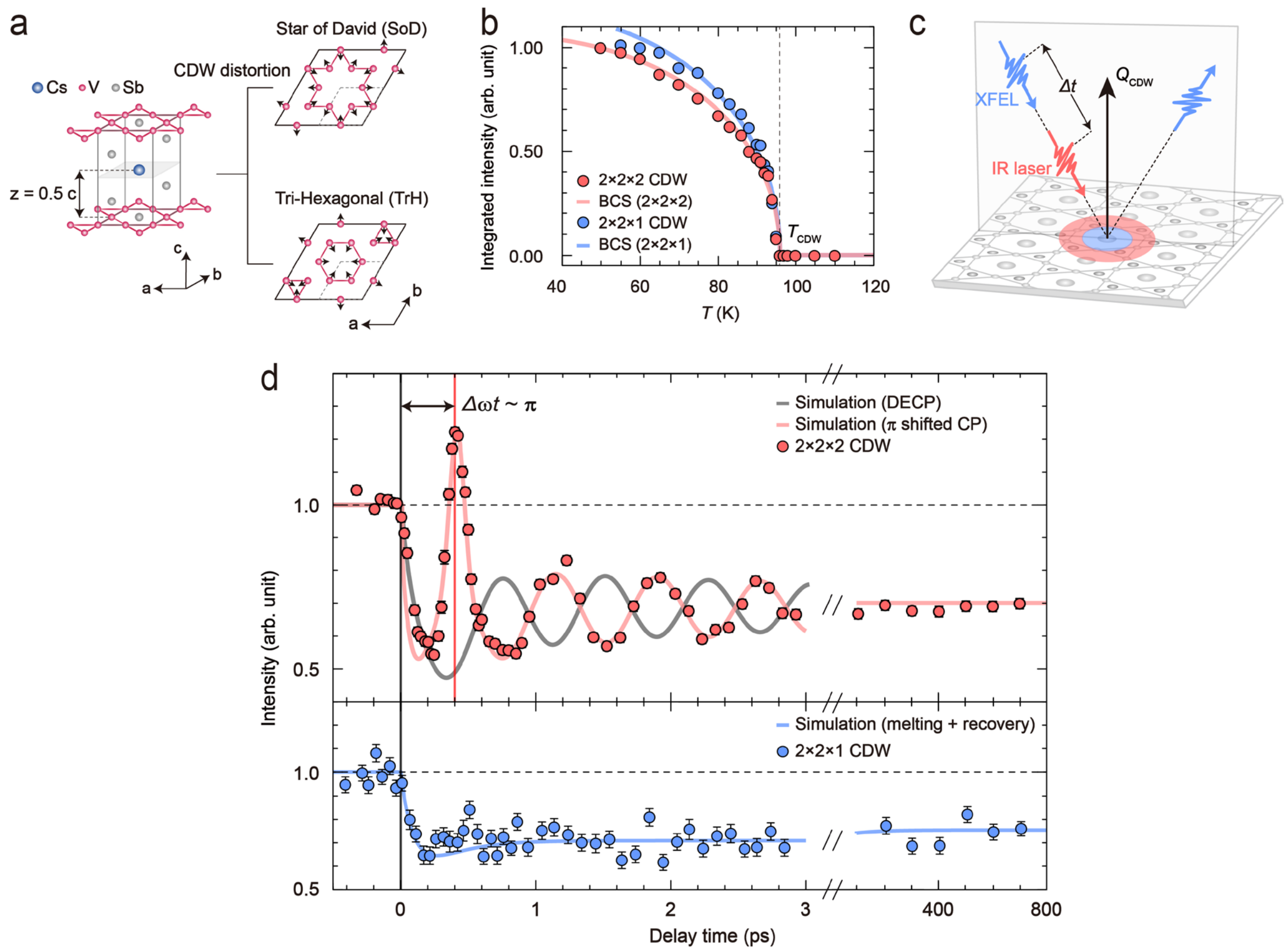

**Fig. 1: Time-resolved X-ray scattering investigation of CDW superstructures of $CsV_3Sb_5$.** **a.** Crystal structure of $CsV_3Sb_5$ with the vanadium-networked Kagome lattice. **b.** CDW order parameters derived from the integrated intensities of the (0.5 0 2) and (0.5 0 1.5) reflections representing the 2×2×1 and 2×2×2 superstructures, respectively. Solid lines depict results from BCS gap equations. **c.** Schematic representation of time-resolved X-ray scattering experiments. **d.** Temporal evolution of the 2×2×1 and 2×2×2 CDW reflections showing photoinduced melting and recovery. Ultrafast melting within 100 fs is commonly observed for both 2×2×1 and 2×2×2, with the peak intensity not fully recovering immediately. Solids lines are from simulations (text). Notably, intensity oscillation is observed for the 2×2×2 CDW. It results from a coherent phonon oscillation (red line) but with a phase delay of $\omega t$ ~ 400 fs, distinguished from normal displacement excited coherent phonon (DECP, grey line)

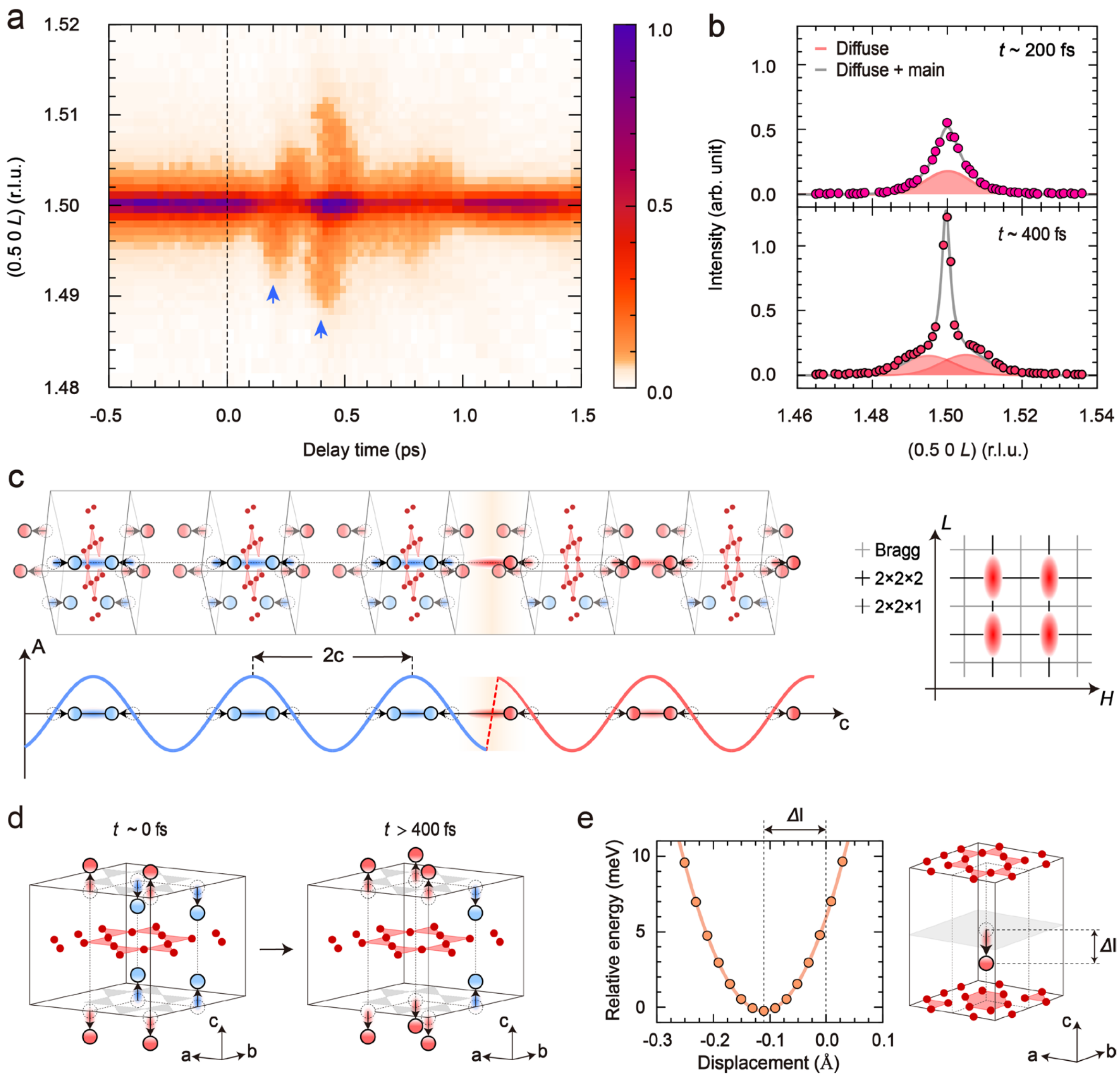

**Fig. 2: Metastable CDW states with localised phonon mode. a.** Photoinduced melting of the 2×2×2 CDW with strong diffuse signals developed along the *c*-direction at ~ 200 fs and 400 fs (blue arrows). **b.** The *L*-scan at 400 fs displays diffusive two-wing structure, centered on 2×2×2 CDW reflection, caused by a phase slip in the CDW domain (lower panel), distinguishable from the signal at 200 fs with diffuse tails resulting from reduced domain size (upper panel). **c.** Phase-slip in the 2×2×2 CDW. The red-coloured sinusoidal wave indicates the domain with three-one Cs displacements newly formed from the native two-two vibration (blue colour). Right panel displays the diffuse scattering pattern corresponding to this phase-slip structure. **d.** Comparison between the symmetric but energetically frustrated two-two vibration in the native CDW and the modified, frustration-relieved Cs vibration initiated at ~ 400 fs, which persists

as a metastable CDW ordering. **e**. Potential energy surface calculation with an eye-guide solid line demonstrates that Cs ions, located at the centre of the 2×2 dimerised V structure, become more stable by moving toward (or out of) the TrH (SoD) mesh.

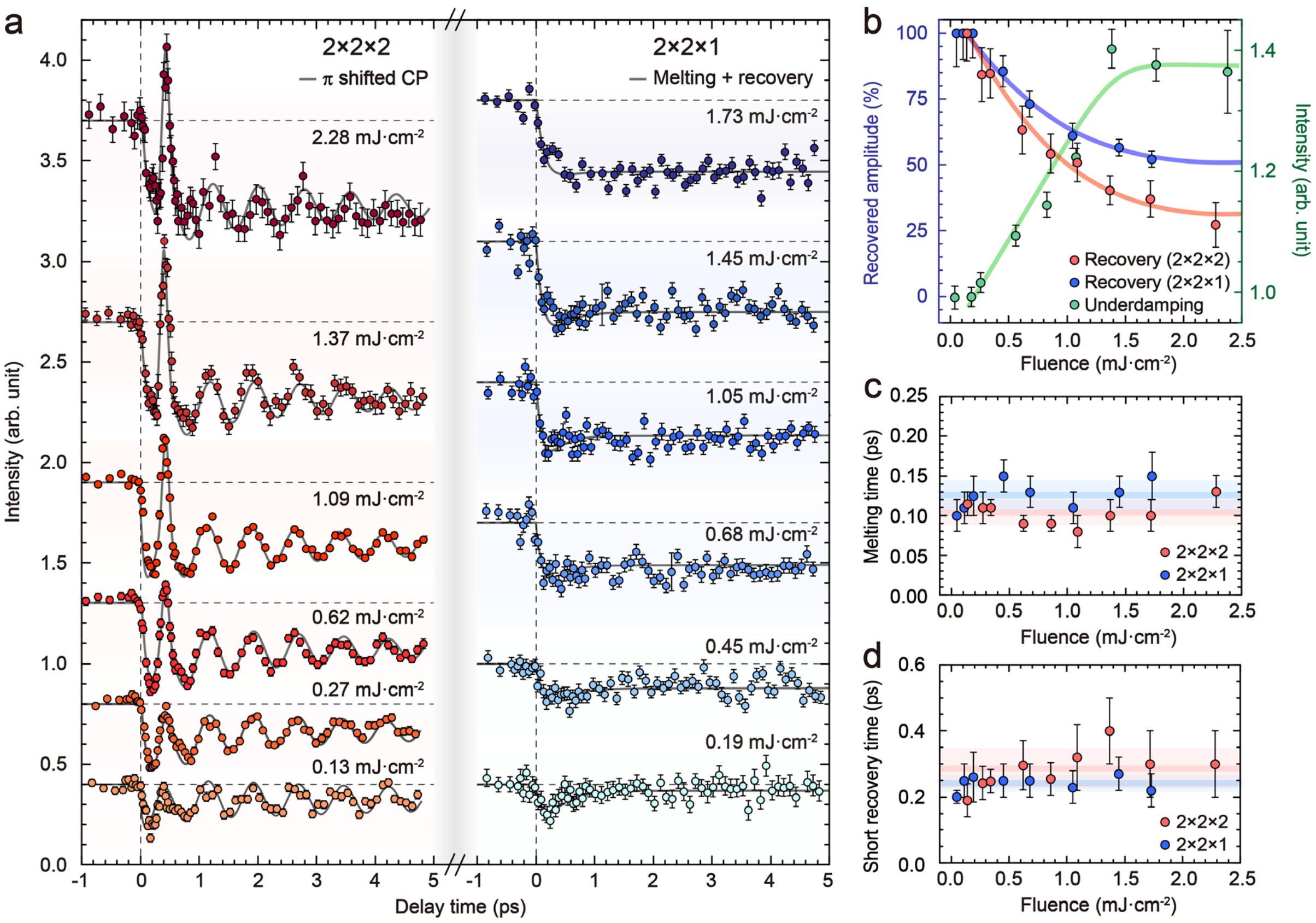

**Fig. 3: Laser fluence dependence of CDW melting and recovery. a**. Laser fluence dependence of the 2×2×2 (left) and 2×2×1 CDW (right) peak intensities, showing a larger reduction in peak intensities with higher laser fluence. Underdamped intensity at ~ 400 fs develops more strongly in the 2×2×2 CDW with increasing laser fluence, but the oscillation period is insensitive to the laser fluence. **b.** Intensities of partially recovered CDWs obtained for different laser fluences, showing a similar behaviour of gradual decrease until ~ 1.4 mJ $cm^{-2}$ before saturation. Intensity of the underdamped oscillation at 400 fs is also compared, displaying a similar pattern of monotonic increase until 1.4 mJ $cm^{-2}$ followed by saturation (right). **c-d.** Comparison of melting and short recovery time of the two CDWs, displaying a similar timescale for melting in 100 fs and short recovery in ~ 300 fs without notable dependence on laser fluence. Solid lines, in b to d, are guide to eyes.

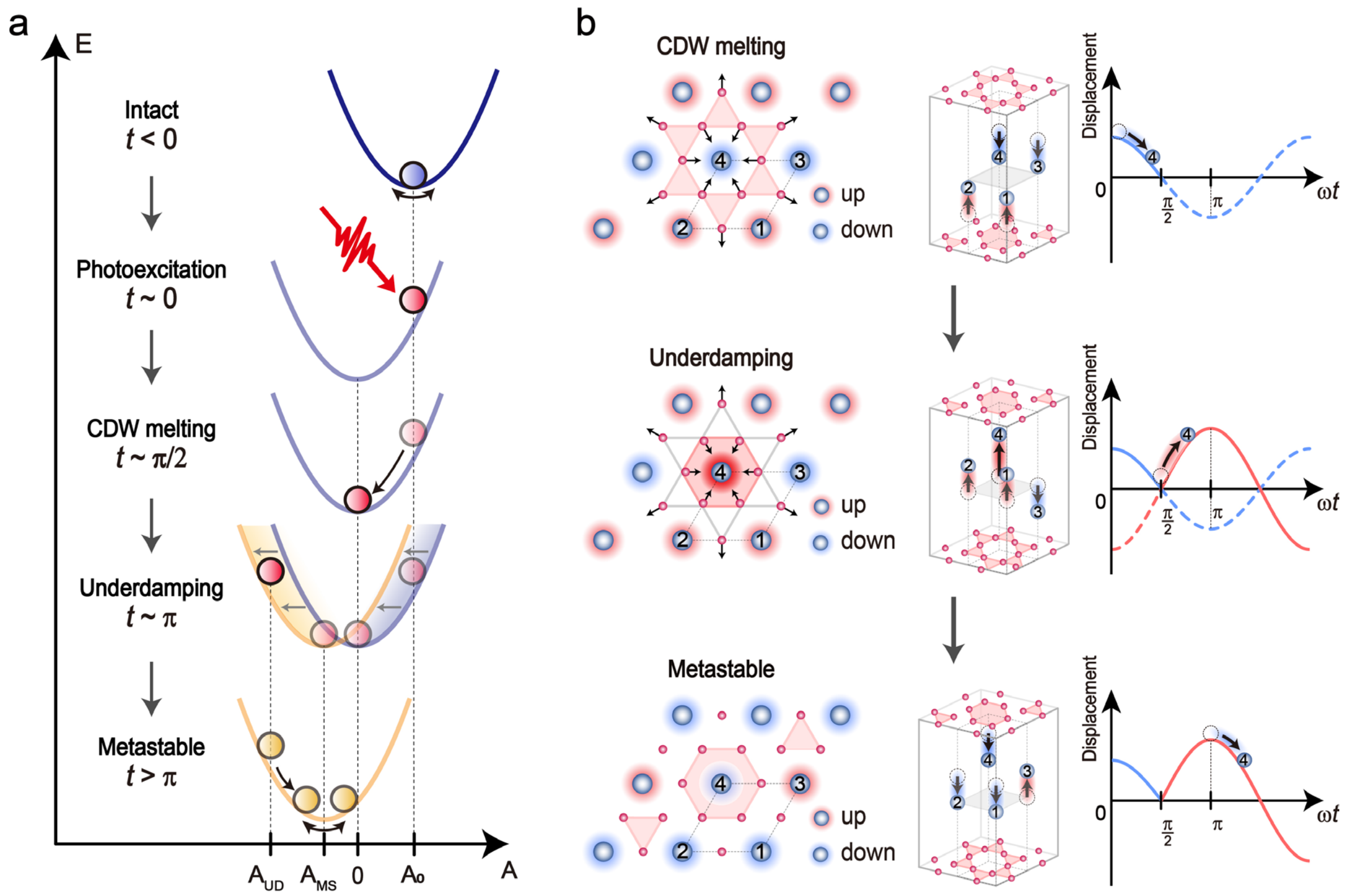

**Fig. 4: Photoinduced underdamping and metastable state of 2×2×2 CDW. a.** Ultrafast reaction in the 2×2×2 CDW depicted by a pictorial view of photoinduced potential energy changes with the CDW amplitude (A). Fs-IR pumping shifts the potential energy to the ground state without CDW (A = 0) caused by photoinduced gap closing ($t$ ~ 0). Following this transiently formed potential energy, the CDW melts to reach the energy ground state ($\omega t \sim \pi/2$). While adapting to this new potential, underdamped displacement of the CDW occurs ($\omega t \sim \pi$) with further displacement of the potential energy surface having the ground state at $A_{MS}$. The CDW amplitude now oscillates around the new ground state amplitude, $A_{MS}$, of the metastable state. **b.** Fs-IR pumping alters Cs ions' $c$-directional vibration ($L_2^-$ phonon) in a manner to relieve energetically frustrated two-two displacements. Upon abrupt disturbance of the electron distribution by fs-IR pumping, the frustrated symmetric (two-up and two-down) phonon mode changes to three-one, compatible with the total energy.

## Methods

### Single crystal growth

Single-crystals of $CsV_3Sb_5$ were grown using typical self-flux methods[11,12]. Owing to the highly reactive nature of elemental Cs, the subsequent preparation of $CsV_3Sb_5$ was conducted inside an argon-filled glove box. Cs liquid (99.98% Alfa Aesar), acid-etched vanadium granules (99.7% Alfa Aesar), and Sb shots (99.99% Alfa Aesar) were placed in alumina crucibles with frit discs, then sealed in Ar-gas purged evacuated quartz tubes. The ampule was heated to 1000 °C for 24 h, then slowly cooled to 600 °C at a cooling rate of 5–3 °C/h. The ampule was then centrifuged to remove the flux, with any remaining flux removed via mechanical cleavage.

### Time-resolved X-ray scattering experiments

Time-resolved X-ray scattering (tr-XRS) experiments were conducted at the resonant soft X-ray scattering end station of the PAL-XFEL[48]. A single-crystal specimen of $CsV_3Sb_5$, with its surface normal parallel to the *c*-axis, was mounted on the cold finger of a liquid helium cryostat (base temperature ~ 20 K). The diffractometer was aligned to have a horizon scattering plane with the sample *ac* plane and π-polarised incoming X-rays. X-ray pulses had a full width at half maximum pulse duration of approximately 80 fs and a repetition rate of 60 Hz. Incident X-ray photon energies were chosen to target the CDW superstructure reflections: 1240 eV for the 2×2×2 CDW at (0.5 0 1.5) reflection and 980 eV for the 2×2×1 CDW at (0.5 0 1). IR photoexcitation of the specimen was induced using a femtosecond Ti:sapphire laser system (wavelength of 800 nm and pulse duration of 50 fs root-mean-square value). The mechanical delay stage controlled the time delay between the pump and probe pulses. The focused X-ray spot size at the sample position was 100 (H) × 200 (V) μm², safely within the fs-IR laser footprint of 500 (H) × 500 (V) μm². Each single-shot X-ray scattering signal was collected using an avalanche photodiode equipped with a high-speed digitiser. For all data collection processes, data were acquired at 30 Hz by interleaving pulses for data without laser pulses to ensure the full recovery of the specimen.

**X-ray scattering measurement with temperature dependence**

The CDW order parameters were determined from the integrated intensity of the CDW reflections through X-ray scattering experiments conducted at the synchrotron 6A beamline of the PLS. Crystals were cryo-cooled using a liquid helium cryostat. The photon energy was fixed at 1580 eV, covering the 2×2×2 (0.5 0 1.5) and 2×2×1 (0.5 0 2) reflections.

**Time-dependent density functional calculation**

To investigate the total energy, optimised lattice parameters, and electronic structure in the ground state of $CsV_3Sb_5$ through first-principles calculations, density functional theory calculations were performed using the Quantum Espresso package[49]. The optimised lattice parameters for the 2×2×2 trigonal $CsV_3Sb_5$ system were $a$ = 10.88 Å and $c$ = 18.66 Å. Kohn-Sham wave functions were illuminated by a plane wave with an energy cut-off of 60 Ry. The Perdew (Burke) Ernzerhof generalised gradient approximation functional was employed to describe electron-electron exchange and correlations[50]. Core electrons and their effects on valence electrons were addressed using norm-conserving pseudopotentials. The Brillouin zone was sampled using a Monkhorst pack with a 3×3×2 mesh. Ionic structure relaxation was achieved by optimising positions with the force criteria of $1.0\times10^{-5}$ eV Å$^{-1}$.

To investigate light-induced dynamics using a first-principles approach, we employed Ehrenfest dynamics with real-time TD-DFT[51]. For time propagation, we utilised a Crank-Nicolson-type time-evolution operator with a time step of 4.8 attoseconds. The optimised Alternative SoD and TrH superstructures[10,21] served as the initial geometries for the TD-DFT calculations. An oscillating electric field of gaussian-packet type (E(t)) was applied to the system through the vector potential term, with $E(t) = dA/dt$ with $A(t) = A_0 e^{-0.5\,[(t-t_0)/\sigma]^2} \sin \omega t$ . The laser parameters in the simulation were set to match experimental conditions: $\hbar\omega = 1.54$ eV and $\sigma = 80$ fs, with a laser power density of $3.6\times10^{10}$ W $cm^{-2}$ and 25% reduced power of $2.7\times10^{10}$ W $cm^{-2}$. Both power densities produced the same results qualitatively supporting that the interpretation is generally valid without specific laser fluence dependence. The detailed calculation settings mirrored those of the DFT calculations, except for the Brillouin zone sampling (1×1×1).

**Data Availability**

The data supporting the findings of this study are available from the corresponding authors upon the request.

**Methods only references**

**Acknowledgements**

C.S. appreciates Prof. K.-W. Kim for insightful discussion on the coherent phonon dynamics and Prof. G.-D. Lee on the abnormal lattice vibrations of interlayer alkali metal ions. X-ray scattering experiments at PAL are approved by the Korean Synchrotron Users Association (KOSUA). This work was supported by the National Research Foundations (NRF) of Korea (grant Nos. 2020K1A3A7A09080405, 2022M3H4A1A04074153, RS-2024-00346711, RS-2024-00333664 and RS-2023-00218180). The computational work was supported by the National Supercomputing Center with supercomputing resources including technical support (KSC-2024-CRE-0124).

**Contributions**

C.S. supervised the project. C.S. and J.-H.P. conceived the project. S.-P.H., H.L., E.P., S.Y.L., J.H., B.L. H.J., and W.-S.N. performed the X-ray scattering experiments. C.W. synthesized the crystals. S.-P.H., H.L., and C.S. analyzed the data. H.C. maintained the fs-IR laser. S.Y.P. worked on the DAQ system. H.K. and D.S. performed the first-principle calculations. S.-P.H. and C.S. wrote the manuscript with input from all authors.

**Corresponding authors**

Correspondence should be addressed to D.S. and C.S.

**Competing interests**

The authors declare no competing interests.

**Supplementary information**

# Frustrated phonon with charge density wave in vanadium Kagome metal

**Authors:**

Seung-Phil Heo, Choongjae Won, Heemin Lee, Hanbyul Kim, Eunyoung Park, Sung Yun Lee, Junha Hwang, Hyeongi Choi, Sang-Youn Park, Byungjune Lee, Hoyoung Jang, Woo-Suk Noh, Jae-Hoon Park, Dongbin Shin*, Changyong Song*

**Supplementary Note 1. Longitudinal scans along the c-axis in coexisting CDW orderings.**

Longitudinal scans along the *(0 0 L)* direction are obtained around the (0.5 0 1.5) and (0.5 0 2) superstructure reflections, representing the 2×2×2 and 2×2×1 CDW, respectively (Fig. S1).

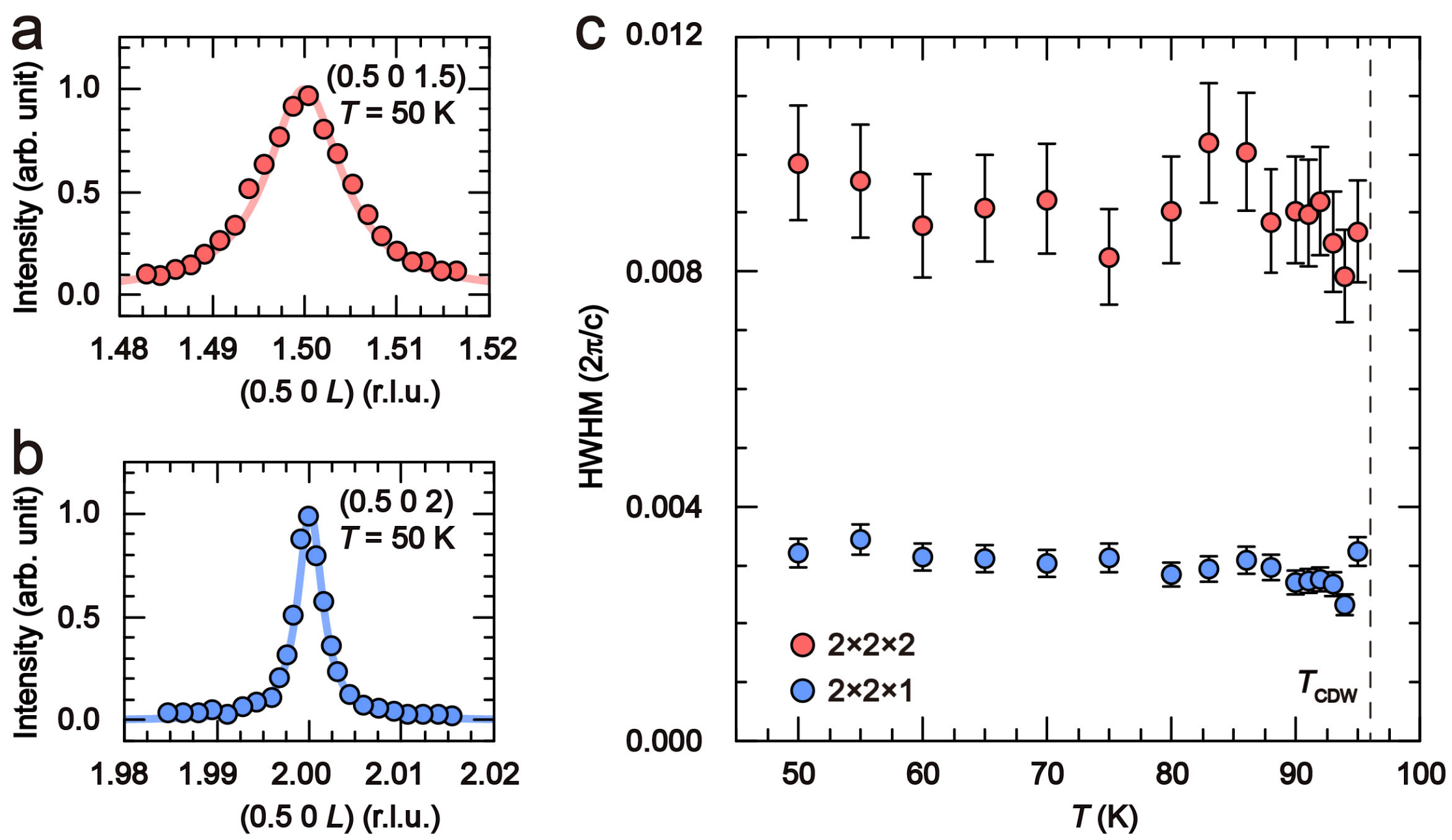

**Figure S1.** *L*-directional scans of 2×2×2 and 2×2×1 CDW. **a-b,** *c*-directional scan around the CDW superstructures of $Q_{CDW}^{2\times2\times2}$ at (0.5 0 1.5) and $Q_{CDW}^{2\times2\times1}$ at (0.5 0 2), data taken at 50 K. **c,** Temperature dependence of the X-ray scattering peak width of the two CDW reflections showing the domain size along the *c*-direction is obtained. Larger peak width is noted for the 2×2×2 CDW implying smaller domain size along the *c*-direction, and no qualitative change in the domain size is noted throughout the temperature range in the CDW phase.

**Supplementary Note 2. Temporal evolution of the photoinduced CDW melting**

Temporal evolutions of the superstructure reflections of the 2×2×1 and 2×2×2 CDWs are quantified by simulating the time dependent intensity variation with the following relations:

$$I_{norm}(t) = 1 - A_{melt}\left(1 - exp\left[-\frac{(t-t_0)}{\tau_{melt}}\right]\right)$$
$$+A_{short}\left(1 - exp\left[-\frac{(t-t_0)}{\tau_{short}}\right]\right) + A_{long}\left(1 - exp\left[-\frac{(t-t_0)}{\tau_{long}}\right]\right).$$

Here, $I_{norm}(t)$, represents the intensity normalized to the intact intensity value, $A_{melt}$ ($\tau_{melt}$ ), $A_{short}$ ($\tau_{short}$ ), $A_{long}$($\tau_{long}$ ) are amplitude (time) of the melting, short recovery and long-time recovery in order. Whilst the 2×2×1 CDW is well described by this simple relaxation function, dynamics of the 2×2×2 CDW involves further complexity due to the coherent phonon oscillation requiring additional terms in simulating the intensity.

The coherent phonon oscillation and underdamped peak are considered with the following relation:

$$I_{norm}(t) = 1 - A_{melt}\left(1 - exp\left[-\frac{(t-t_0)}{\tau_{melt}}\right]\right)$$
$$+A_{short}\left(1 - \exp\left[-\frac{(\mathrm{t}-\mathrm{t}_0)}{\tau_{\mathrm{short}}}\right]\right) + A_{long}\left(1 - exp\left[-\frac{(t-t_0)}{\tau_{long}}\right]\right)$$
$$+H(t-t_0) * A_{osc}\exp\left[-\frac{(t-t_0)}{\tau_{osc}}\right] * \cos[2\pi f(t-t_0)+\pi]$$
$$+A_{UD}\exp\left[-\frac{1}{2}\frac{(\mathrm{t}-\tau_{UD})^2}{\sigma^2}\right].$$

Here, $A_{osc}$ ($\tau_{osc}$ ) and $A_{UD}$ ($\tau_{UD}$ ) are amplitude (characteristic time) of the coherent phonon oscillation and underdamped CDW recovery, respectively. Phonon frequency is noted by $f$. It should be noted that a phase delay of π in coherent phonon oscillation is essential to interpret intensity to launch the delayed phonon at ~ 400 fs ($2\pi \times 1.3$ THz $\times\ 400\ fs\ \sim \pi$) (Fig. 1d).

**Supplementary Note 3. Temporal evolution of the Bragg reflection.**

The temporal evolution of the (0 0 1) Bragg reflection intensity is displayed in Fig. S2, where we used the following function to fit the data:

$$I_{norm}(t) = 1 - A_{melt}\left(1 - exp\left[-\frac{(t-t_0)}{\tau_{melt}}\right]\right) + A_{rec}\left(1 - exp\left[-\frac{(t-t_0)}{\tau_{rec}}\right]\right).$$

In the case of Bragg reflection, the melting time of $\tau_{melt}$ ranges between 50 and 60 ps consistent with ordinary lattice melting time scale[1]. The incident laser fluence is approximately 1.6 mJ cm$^{-2}$.

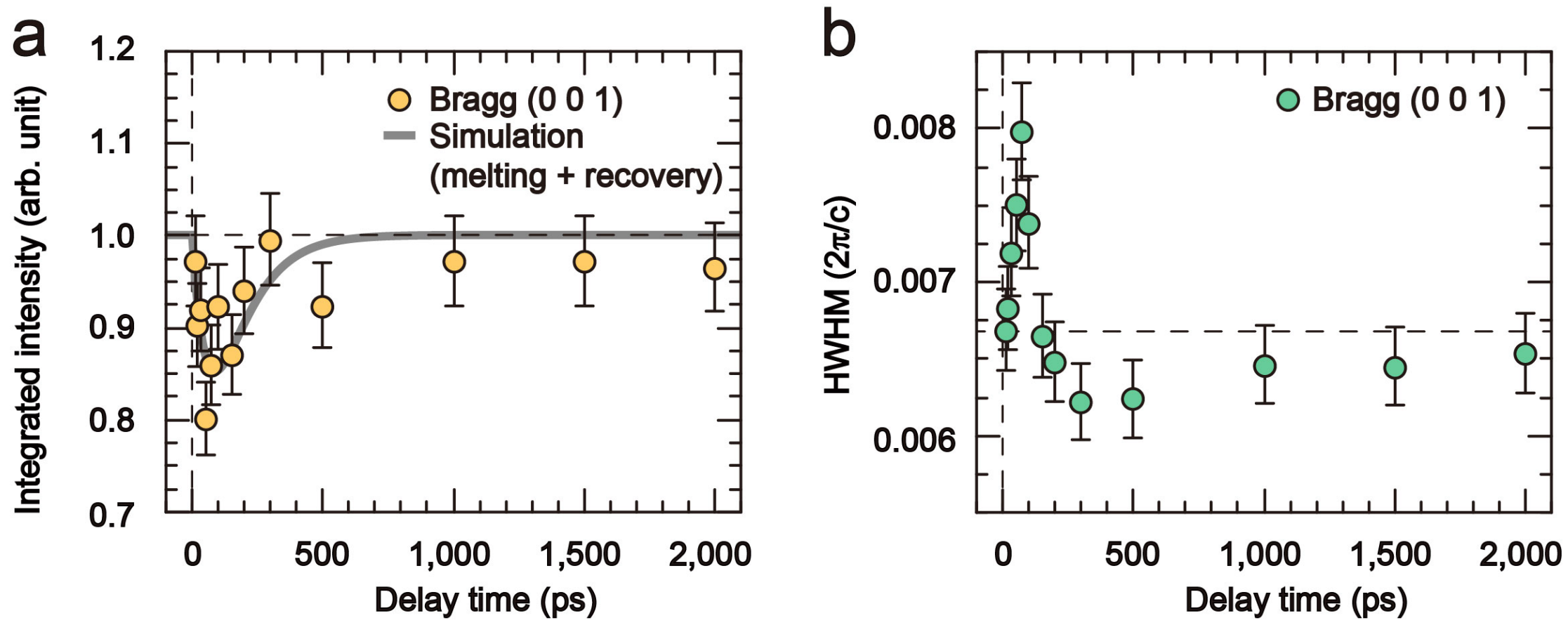

**Figure S2.** Temporal evolution of the (0 0 1) Bragg reflection. **a,** Reduction in the integrated intensity in the time scale of several tens of ps and recovery in several hundred ps to imply that thermalized lattice recovers before 1 ns. The simulated function (see the text) is drawn with the solid gray line. **b,** Half width at half-maximum of the (0 0 1) Bragg reflection shown in reciprocal lattice unit.

**Supplementary Note 4. X-ray structure factor considering the $L_2^-$ coherent phonon at $\boldsymbol{Q} = \boldsymbol{Q}_{\mathbf{CDW}}^{2\times2\times2}$**

X-ray scattering intensity for a given momentum transfer, $\boldsymbol{Q}$, is obtained using $I(Q) \sim |F(Q)|^2$, with the $F(Q)$ representing the structure factor. In general, the structure factor can be acquired by:

$$F(\boldsymbol{Q}) = \underbrace{\sum_n exp[i\boldsymbol{Q}\cdot R_n]}_{lattice}\underbrace{\sum_j f_j(\boldsymbol{Q})\, exp[i\boldsymbol{Q}\cdot \boldsymbol{r}_j]}_{unit\ cell} = \delta_{\boldsymbol{Q}=\boldsymbol{G}}\sum_j f_j(Q)\, exp[i\boldsymbol{Q}\cdot \boldsymbol{r}_j]\ .$$

Here, $\boldsymbol{R}_n$ and $\boldsymbol{G}$ are real and reciprocal lattice vectors, respectively. The atomic form factor, spatial Fourier transformation of the charge distribution, is expressed as $f_j$ and $r_j$ denotes the $j$th atom in the unit cell. With this, any change in the X-ray structure factor of the 2×2×2 superstructure caused by the $L_2^-$ phonon can be obtained using,

$$\Delta F(\boldsymbol{Q}) \sim \frac{dF(\boldsymbol{Q})}{d\boldsymbol{r_j}}\Delta\boldsymbol{r_j} = \delta_{\boldsymbol{Q}=\boldsymbol{G}}\sum_j f_{Cs}(\boldsymbol{Q})\, exp[i\boldsymbol{Q}\cdot \boldsymbol{r}_j^{\mathbf{0}}]\Delta\boldsymbol{r_j}$$

Here, the vibration of Cs atoms is considered to account for the $L_2^-$ phonon exclusively. The displacement of Cs ions associated with $L_2^-$ phonon is denoted as $\Delta\boldsymbol{r_j}$, while their initial positions in the 2×2×2 unit cell are represented as $\boldsymbol{r_j^0}$. Please note that we have used 2×2×2 superstructure as a unit cell in calculating this structure factor. Here, $(2n + 1\ 0\ 2n + 1)$ is equivalent to $(l + 0.5\ 0\ l + 0.5)$, with $n$ and $l$ integers, in normal 1×1×1 unit cell.

The possible vibration modes of the $L_2^-$ phonon along the $3\boldsymbol{Q}$ symmetry—corresponding to $\boldsymbol{Q}_1$ = (0.5 0 0.5), $\boldsymbol{Q}_2$ = (0 0.5 0.5), and $\boldsymbol{Q}_3$ = (-0.5 -0.5 0.5) —are depicted in Fig.S3. The structure factor modification occurs only when the $\boldsymbol{Q}$ position matches the symmetry of the phonon mode; otherwise, no modification is observed.

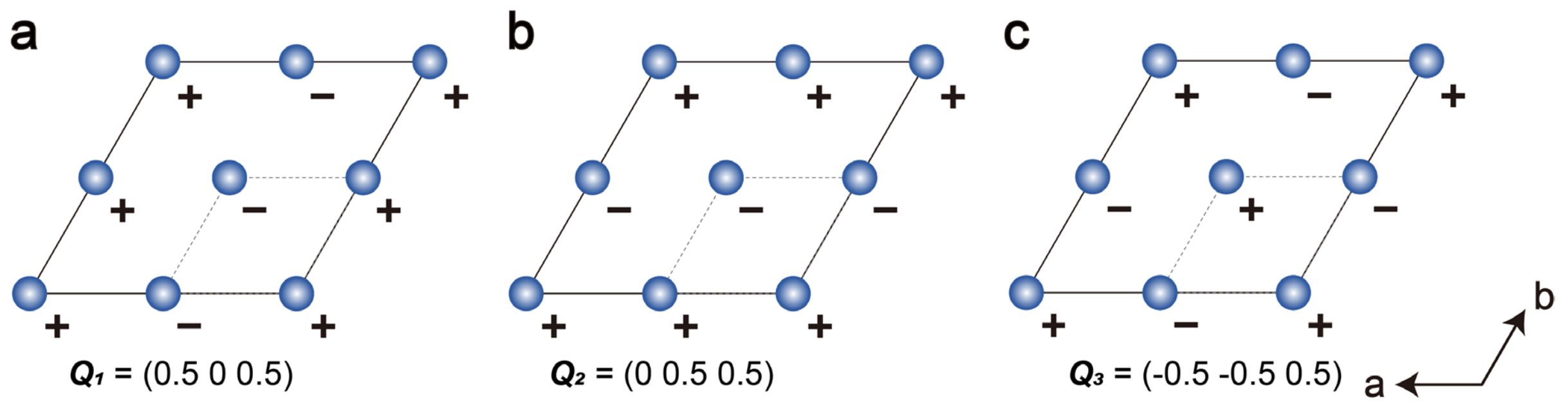

**Figure S3.** Possible $L_2^-$ phonon modes, which share the same symmetry but differ by a 120° rotation, corresponding to degenerate 3$\boldsymbol{Q}$ positions: $\boldsymbol{Q_1}$ = (0.5 0 0.5) (a), $\boldsymbol{Q_2}$ = (0 0.5 0.5), and $\boldsymbol{Q_3}$ = (-0.5 -0.5 0.5). The plus and minus signs indicate the direction of the Cs ion's vibration along the *c*-axis. Here, we only depict the vibration mode in a single layer; the vibration mode in the upper or lower layer is in the opposite direction. From this, only $Q_1$ case contributes to the intensity at (0.5 0 0.5) reflection, and others are vanishing.

**Supplementary Note 5. Phase slip simulation for the diffuse peak near $\boldsymbol{Q} = \boldsymbol{Q_{CDW}^{2\times2\times2}}$**

We constructed a simplified one-dimensional chain model of Cs ions along the *c*-axis as depicted in Fig. 2c. In this model, the chain consists of alternating domains with original and phase-slipped CDW region, introducing a phase difference at the boundary:

$$F(Q) \sim \underbrace{\sum_{n=0}^{N-1} \sum_j exp[iQ\cdot r_n^0] \sum_j exp\left[iQ\cdot r_{n,j}^0\right]}_{Original} + \underbrace{\sum_{m=N}^{N+M-1} exp[iQ\cdot r_m^0] \sum_j exp\left[iQ\cdot r_{m,j}^0\right]}_{Slipped\ phase}$$

with positions defined as:

$$r \begin{cases} r_n^0 = (c*n)\hat{z}, \quad r_{n,j}^0 = A_0 cos(Q\cdot r_n^0) : \ For\ the\ orginal\ chain \\ r_m^0 = (c*m)\hat{z}, \quad r_{m,j}^0 = A_1 cos(Q\cdot r_m^0 - \varphi_S(t)) : For\ the\ slipped\ chain \end{cases}$$

Using this, one can see that the CDW intensity appears at $Q$ = (0 0 0.5) with diffuse satellite relative to the CDW superstructure reflection at (0 0 0.5). The sharp CDW reflection vanishes only under the artificial condition of an exact population of the original and slipped domains, with identical amplitudes for the slipped CDW and a phase difference of precisely 180 degrees.

Further to help understanding the scattering signal, we simulate the scattered intensity observed and interpreted in our manuscript with (Gaussian) random population of the phase slipped domains typically observed in the discommensurate region with the results shown below. This simulation demonstrates the emergence of double-sided broad peaks at $\phi(t) = \pi$, aligning with our observation that these peaks appear around $t \sim 400$ fs.

We note that further increased disorder can also result in the merging of two satellite diffuse peaks into a single one centered at commensurate position. To assess this, we compared the two-wing diffuse fitting with a single-peak fitting scenario, as shown in Figure S5. This comparison reveals that while a single broad peak can be attempted, it fails to capture the intensity at the side wings, confirming that the two-wing diffuse fitting better represents the observed data.

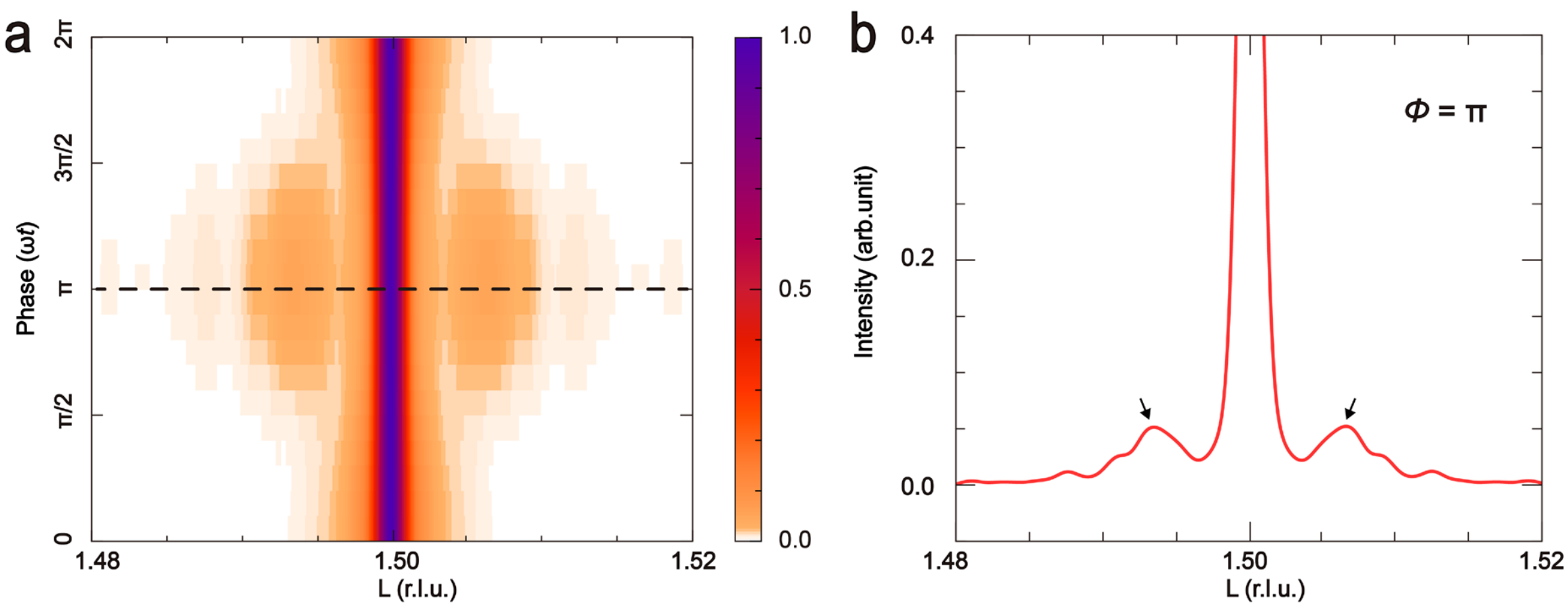

**Figure S4.** Fourier-transformed data from the 1D chain model, shown according to different phase values. **a,** Color map of intensity as a function of phase. As $\phi(t)$ approaches π, two broad diffuse peaks appear alongside the main peak. **b,** the two-wing diffuse peaks emerges, when, $\phi(t) = \pi$, reflecting the occurrence of phase slip at this value.

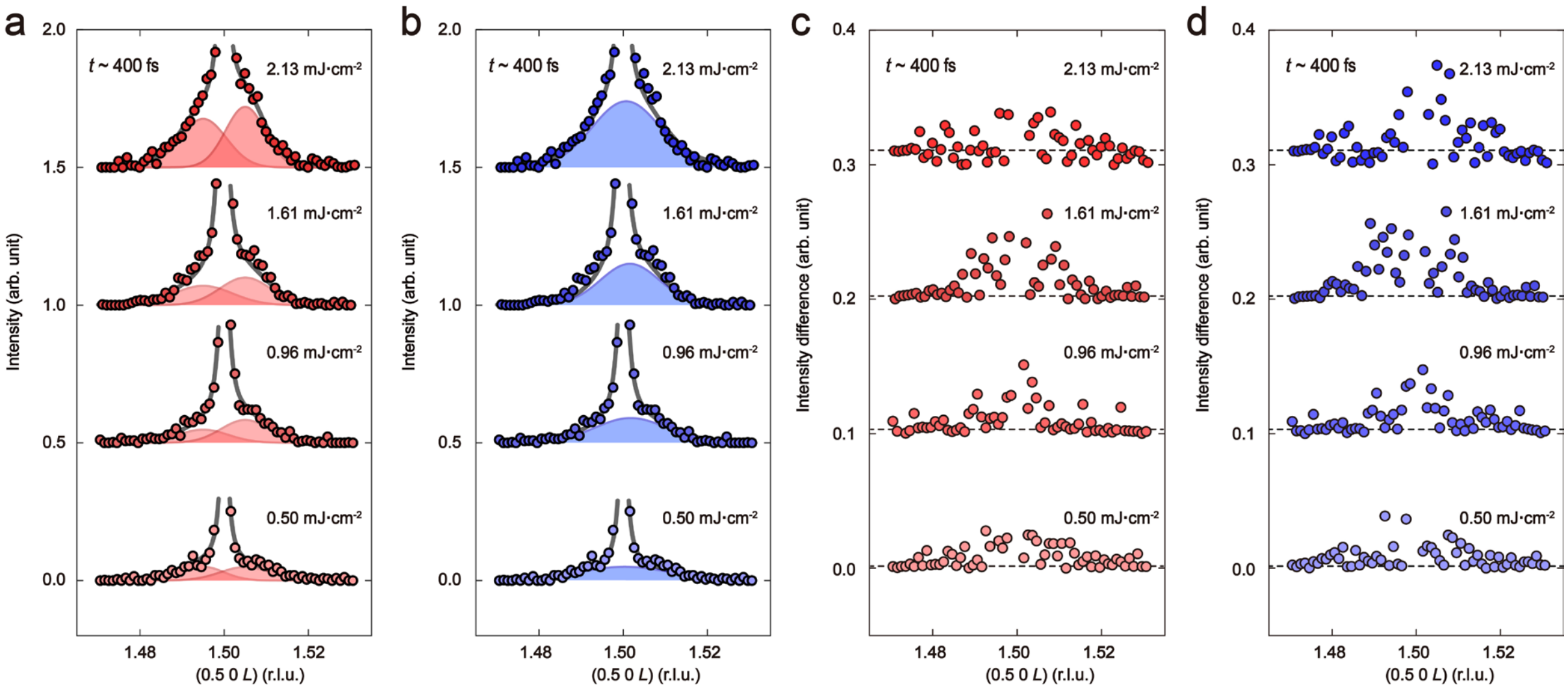

**Figure S5. a-b,** Comparison between the two-wing diffuse fitting (red) (a) and single diffuse fitting (blue) (b) relative to the main CDW reflection at $Q$ = (0.5 0 L). **c-d,** Intensity difference between the fitting function and the dataset. The single-peak fitting (d) shows greater deviation compared to the two-peak fitting (red) (c), particularly in the side regions. Dashed lines indicate the average value of the tail regions in each dataset.

**Supplementary Note 6. Time-dependent DFT (TD-DFT) simulation results**

We performed static and time-dependent density functional theory (TD-DFT) calculation to investigate the light-induced dynamics in $CsV_3Sb_5$. Static DFT calculation shows the dominant contributions of V and Sb orbitals near the Fermi level, showing a significantly higher density of states (DoS) for V and Sb orbitals compared to Cs orbitals (Fig. S6a). TD-DFT calculation (Fig. S6b) shows light-driven excitation of occupied states near the Fermi level, primarily between −2 and 1.5 eV, where excited carriers accumulate at approximately 0.4 eV and 1.2 eV after photoexcitation. This suggests that the laser selectively excites V and Sb orbitals, indirectly affecting Cs ions through the CDW interaction. We note that the distinction between the static density of state and TD-DFT comes from the method involved heavy computational load. Static DOS is from 9×9×3 *k*-point sampling (Fig. S6a) and time-dependent analysis (Fig. S6b) is obtained from a gamma *k*-point sampling. However, one can find that they both deliver the same message in describing behavior on light-excited V and Sb orbital states.

The temporal evolution of Cs ion's motion shows the corresponding 3-1 Cs displacement and velocity (Fig. S6c and Supplementary Video 1.) due to photoexcitation as an experimental observation.

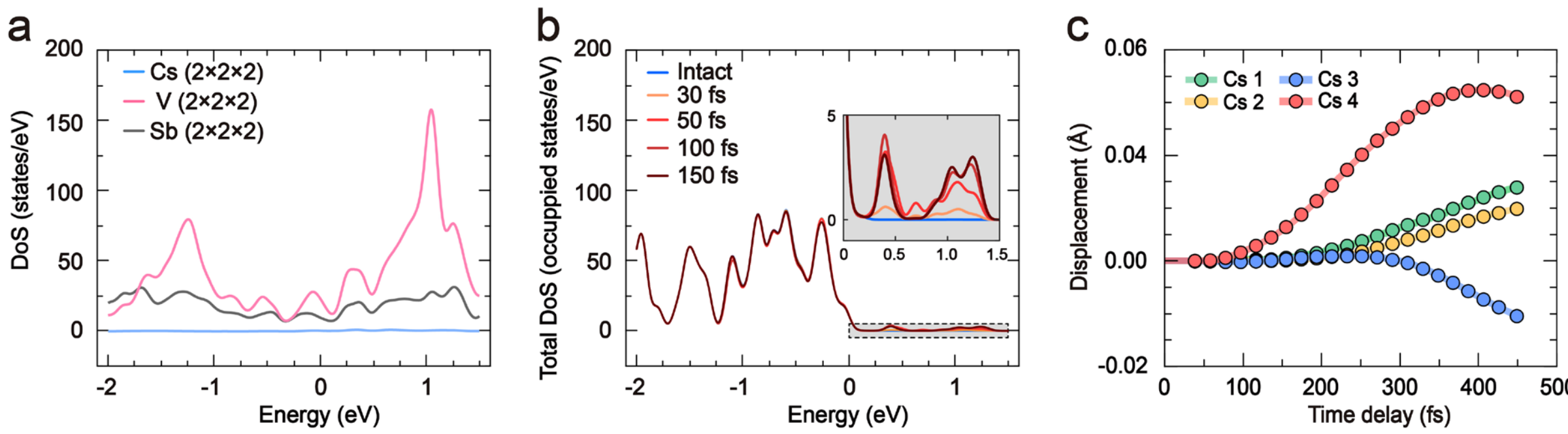

**Figure S6. a,** The projected density of states near the Fermi level (-2 to 1.5 eV) with 2×2×2 CDW formation, indicating dominant contributions from V and Sb orbitals. **b,** The time-

varying occupied states from TD-DFT, showing changes in the V and Sb orbitals over a selected time delay (0 to 150 fs), confirming the effect of photoexcitation on these orbitals. **c,** Cs ion's dynamics from TD-DFT. Each atom is labeled with Cs 1, Cs 2, Cs 3, and Cs 4, positioned at coordinates (0 0 0.5), (0.5 0 0.5), (0 0.5 0.5), and (0.5 0.5 0.5), respectively, within the 2×2×2 unit cell in reciprocal lattice unit.

**Supplementary Note 7. Fluence dependence of coherent phonon frequency in 2×2×2 CDW ordering.**

The fluence dependence of modified $L_2^-$ coherent phonon frequency is displayed in Fig. S7. Photoinduced softening in the phonon frequency is noted less than 2.5%.

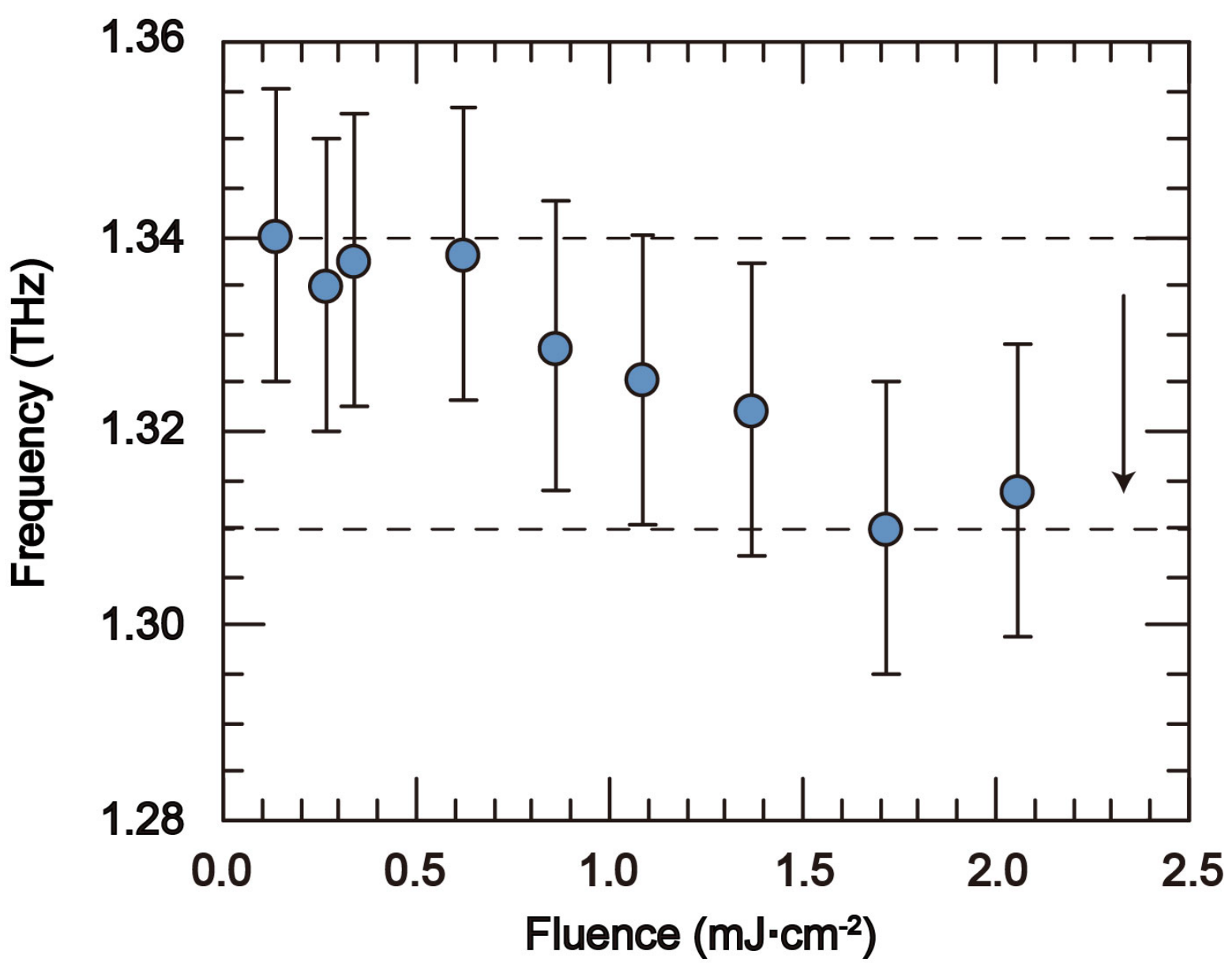

**Figure S7.** Fluence dependence of coherent phonon frequency. The frequency values are extracted from simulated function for temporal evolution of 2×2×2 CDW intensity depicted in supplementary note 2. The upper and lower dot lines represent the maximum and minimum values from all data set. Very weak (2.5%) softening in the phonon frequency is noted.

**Supplementary Note 8. Selected *H* and *L* scans after the photoexcitation**

The longitudinal scans along the ($H$ 0 0) and (0 0 $L$) direction are obtained around the (0.5 0 1) and (0.5 0 1.5) superstructure reflections representing the 2×2×1 and 2×2×2 CDW, respectively (Figs. S8 and S9). The $H$ and $L$ scan at $Q = Q_{CDW}^{2\times2\times1}$ at 200 and 400 fs are similar in its shape to the intact CDW reflection except for decreased intensity. It indicates the intensity reduction results from reduced 2×2×1 CDW amplitude. In contrast, peak broadening appears significantly for the 2×2×2 CDW, and this diffuse peak can be attributed to Cs ions consistent with the independence of the 2×2×1 CDW to Cs ions' displacement. In particular, the reflection at 400 fs consists of three peaks with the main peak at 1.5 having the width same as that of the intact structure, which immediately rules out the scenario of 2×2×2 CDW domain size increase by reducing the 2×2×1 CDW. Two side lopes developed symmetric to the center peak at 1.5 are from the phase slip in 2×2×2 CDW as described in the text.

In all scans, we did not observe any change in domain size with the photoexcitation along the *ab*-plane direction.

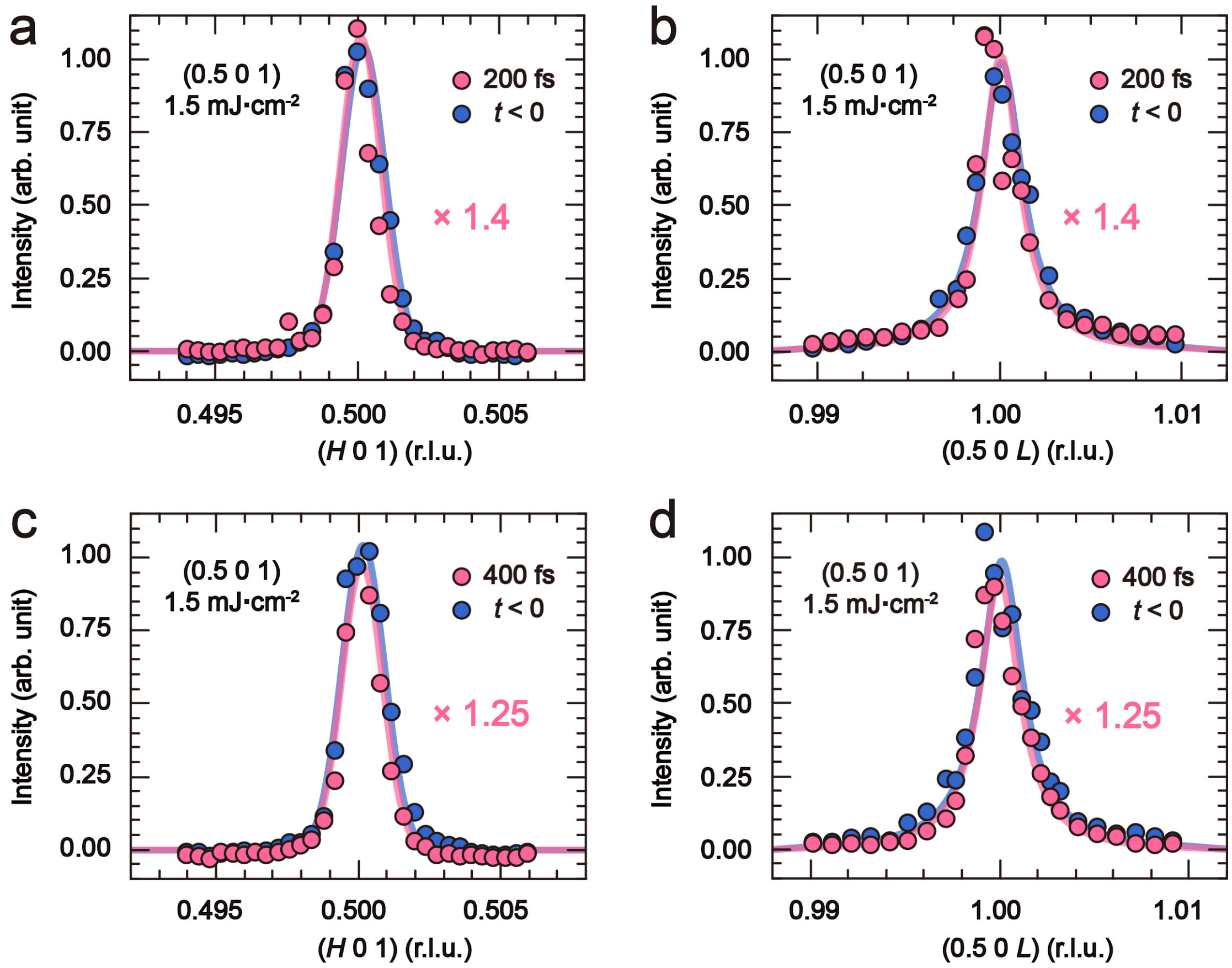

**Figure S8.** *H* and *L* directional scans along the 2×2×1 CDW reflection. Scans performed at 200 fs and 400 fs when the X-ray scattering intensity of the CDW reached minimum and quick recovery to have underdamping for 2×2×2 CDW, respectively.

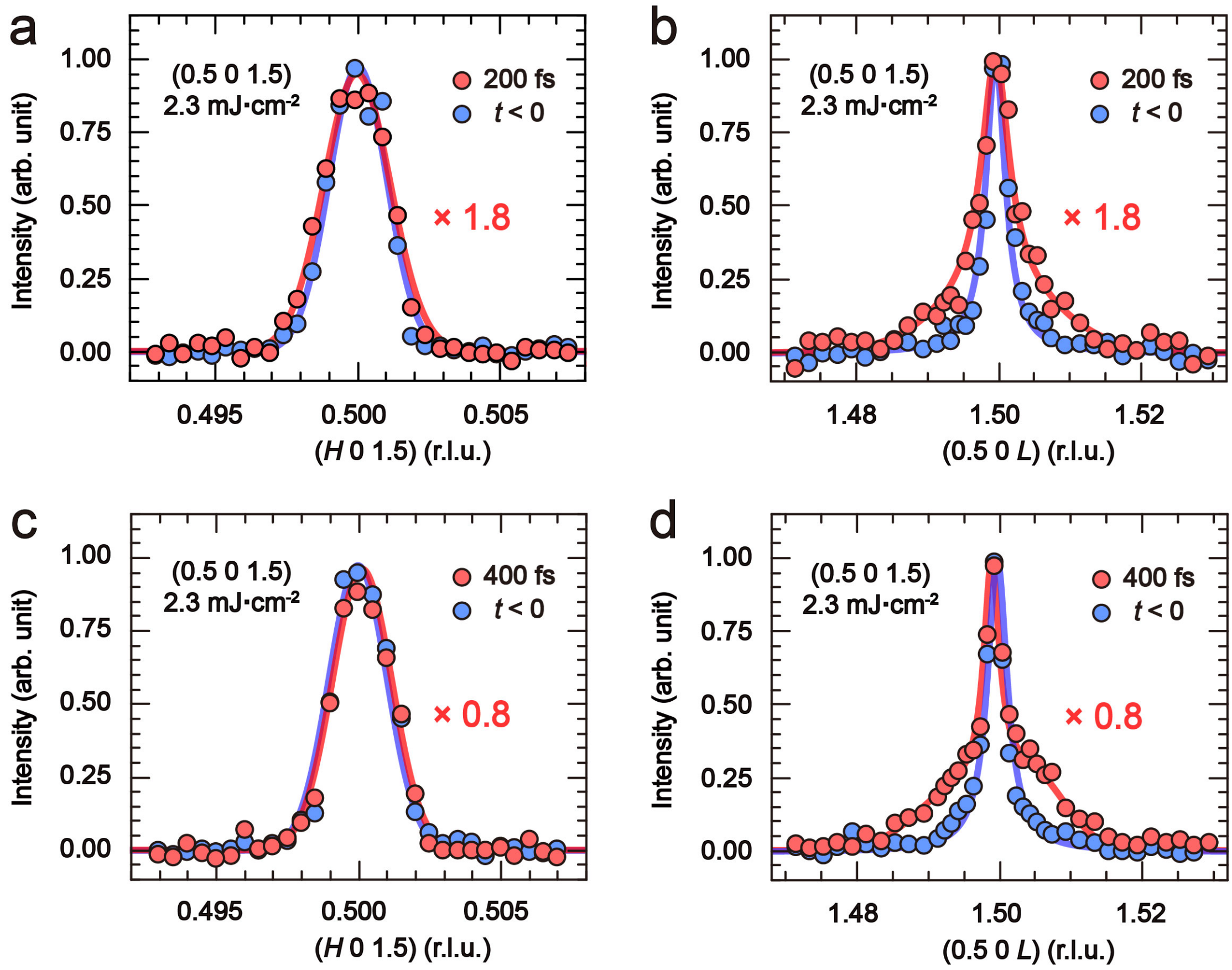

**Figure S9.** *H* and *L* directional scans along the 2×2×2 CDW reflection. Scans performed at 200 fs and 400 fs when the X-ray scattering intensity of the CDW reached minimum and quick recovery to have underdamping for 2×2×2 CDW, respectively. Upon photoinduced CDW melting, the X-ray scattering peak width increases in (b) informing the decreased domain size along the *c*-direction. A three-peak structure with two side lopes along the central peak is noted in (d) with the development of the phase slip.

**Supplementary Note 9. IR-laser fluence dependence of *L* scan at 400 fs.**

The laser fluence dependence of the X-ray reflection from the 2×2×2 CDW along the *L* direction are compared for the scan taken at 400 fs (Fig. S10). Increase in both of the main peak and the diffusive side lopes intensities are noted as the laser fluence increase, but the peak shape remains same.

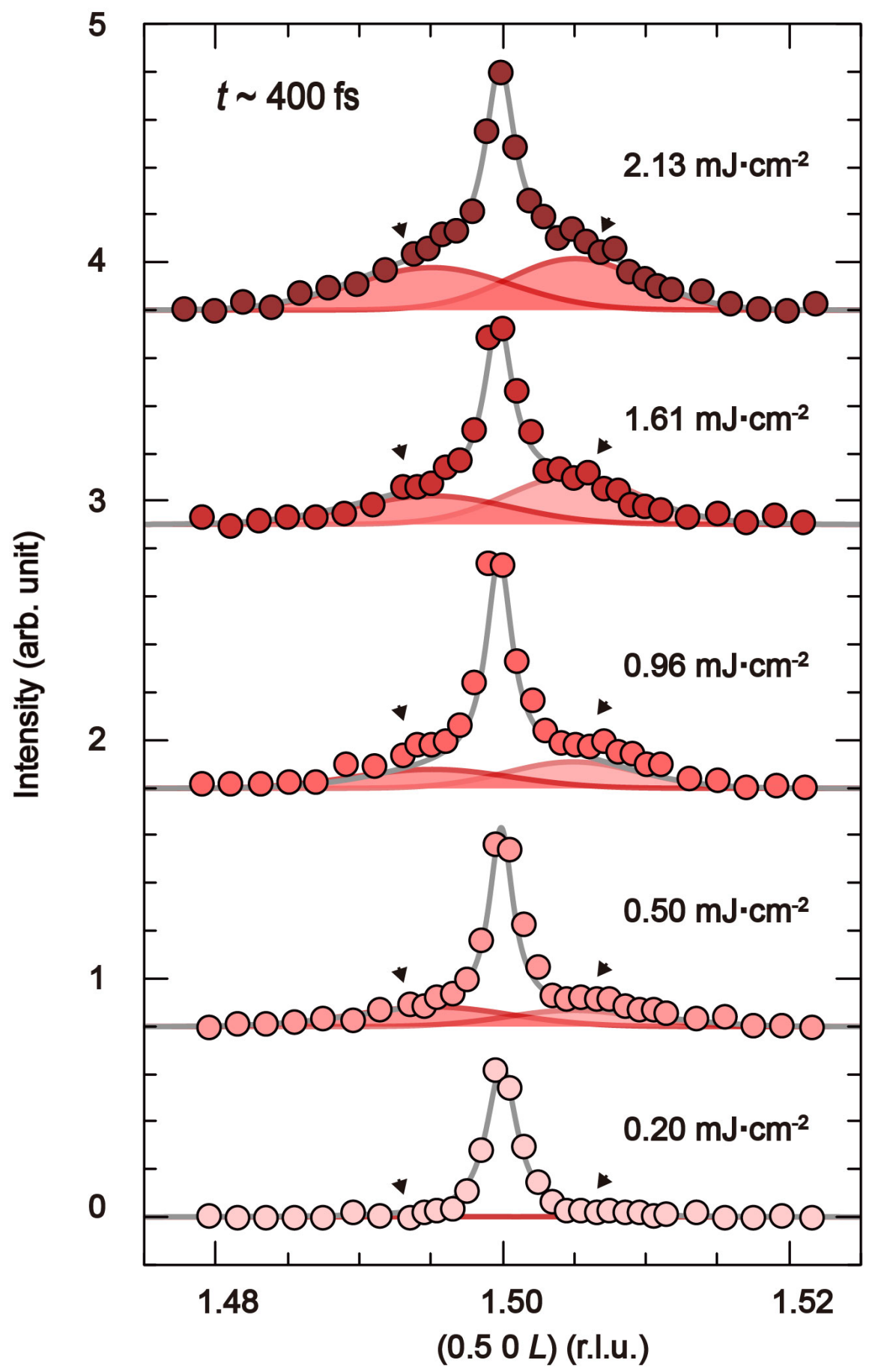

**Figure S10.** Fluence dependence of *L* scan along the (0.5 0 1.5) reflection from 2×2×2 CDW at 400 fs. Experimental data are fit to three Gaussians (grey) of one at the center representing the main CDW reflection and two diffuse reflections (red lines) symmetrically shifted from the center one. The diffusive side lopes are marked with arrow heads.

**Supplementary Note 10. X-ray structure factor of the 2×2×2 CDW at $\boldsymbol{Q} = \boldsymbol{Q}_{CDW}^{2\times2\times1}$**

We calculate the X-ray structure factor considering the different configuration of the vanadium dimerization in the CDW phase. The calculation is performed using the vanadium ions' in-plane displacements listed in the Table S1. Various CDW configurations are considered as shown in Fig. S11[6,7].

| Atom | $x/2a$ | $y/2b$ |
|---|---|---|
| $\boldsymbol{V_1}$ | $0.25 \pm \delta_1$ | $0 \pm 2\delta_1$ |
| $\boldsymbol{V_2}$ | $0 \pm 2\delta_1$ | $0.25 \pm \delta_1$ |
| $\boldsymbol{V_3}$ | $0.25 \mp \delta_2$ | $0.25 \mp \delta_2$ |
| $\boldsymbol{V_4}$ | 0.5 | $0.25 \mp \delta_2$ |
| $\boldsymbol{V_5}$ | $0.25 \mp \delta_2$ | 0.5 |
| $\boldsymbol{V_6}$ | $0.75 \mp \delta_1$ | $0 \mp 2\delta_1$ |
| $\boldsymbol{V_7}$ | $0.75 \mp \delta_1$ | $0.25 \pm \delta_1$ |
| $\boldsymbol{V_8}$ | $0.75 \pm \delta_2$ | 0.5 |
| $\boldsymbol{V_9}$ | $0 \mp 2\delta_1$ | $0.75 \mp \delta_1$ |
| $\boldsymbol{V_{10}}$ | $0.25 \pm \delta_1$ | $0.75 \mp \delta_1$ |
| $\boldsymbol{V_{11}}$ | 0.5 | $0.75 \pm \delta_2$ |
| $\boldsymbol{V_{12}}$ | $0.75 \pm \delta_2$ | $0.75 \pm \delta_2$ |

**TABLE S1.** Atomic positions of vanadium in 2×2 SOD and TrH dimerization

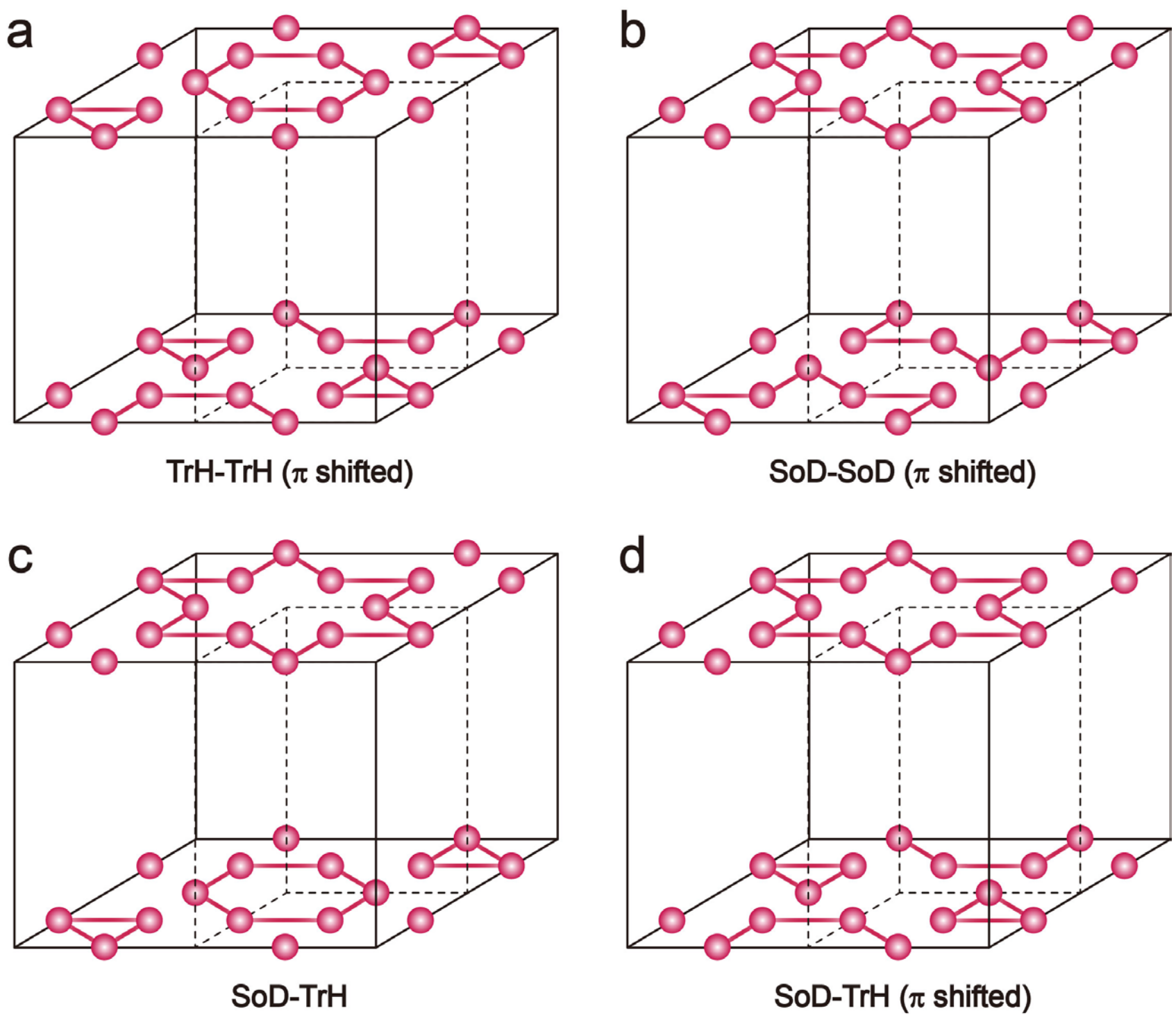

**Figure S11. a-d,** All possible CDW configurations making the 2×2×2 superstructure are shown. **a,** TrH-TrH (π-shifted) **b,** SoD-SoD (π-shifted) **c,** SoD-TrH (3), **d,** SoD-TrH (π-shifted).

The X-ray structure factor for each CDW configuration can then be obtained as:

$$F(Q) = \delta_{Q=G} \sum_j f_V(Q)[\underbrace{exp[iQ \cdot r_j^{TrH}]}_{TrH} + \underbrace{exp[iQ \cdot (r_j^{TrH} + [a_0\ 0\ c_0])]}_{\pi-\text{shifted } TrH}]$$

$$F(Q) = \delta_{Q=G} \sum_j f_V(Q)[\underbrace{exp[iQ \cdot r_j^{SoD}]}_{SoD} + \underbrace{exp[iQ \cdot (r_j^{SoD} + [a_0\ 0\ c_0])]}_{\pi-\text{shifted } SoD}]$$

$$F(Q) = \delta_{Q=G} \sum_j f_V(Q)[\underbrace{exp[iQ \cdot r_j^{SoD}]}_{SoD} + \underbrace{exp[iQ \cdot (r_j^{TrH} + [0\ 0\ c_0])]}_{TrH}]$$

$$F(Q) = \delta_{Q=G} \sum_j f_V(Q)[\underbrace{exp[iQ \cdot r_j^{SoD}]}_{SoD} + \underbrace{exp[iQ \cdot (r_j^{TrH} + [a_0\ 0\ c_0])]}_{\pi-\text{shifted } TrH}].$$

Considering only first order phase term, $e^x \simeq 1 + x$ under condition of $\boldsymbol{Q} \cdot \boldsymbol{\delta}_{1,2} << 1$, the structure factor $F(Q)$ at $Q = (l \pm 0.5\ 0\ n)$ corresponding to the $Q = Q_{CDW}^{2\times2\times1}$ is zero ($l$ and $n$ is integer). It confirms that the possibility of observing a 2×2×2 superstructure domain at $Q = Q_{CDW}^{2\times2\times1}$ can be excluded.